\documentclass[10pt,conference,a4paper]{IEEEtran}
\usepackage{ifpdf}
\usepackage{cite}
\ifCLASSINFOpdf
\usepackage[pdftex]{graphicx}
\DeclareGraphicsExtensions{.pdf,.jpeg,.png}
\else
\usepackage[dvips]{graphicx}
\DeclareGraphicsExtensions{.eps}
\fi
\usepackage{amsmath}
\usepackage{algorithmic}
\usepackage{array}
\ifCLASSOPTIONcompsoc
\usepackage[caption=false,font=normalsize,labelfont=sf,textfont=sf]{subfig}
\else
\usepackage[caption=false,font=footnotesize]{subfig}
\fi
\usepackage{fixltx2e}
\usepackage{url}
\usepackage{bm}
\usepackage{textcomp}
\usepackage{caption2}
\usepackage{multirow}
\usepackage{booktabs}
\usepackage{mathrsfs}
\usepackage{amssymb}
\usepackage{float}
\usepackage{bm}
\usepackage[colorlinks,linkcolor=blue,anchorcolor=green,citecolor=red]{hyperref}
\makeatletter
\begin{document}
	\title{An Efficient Approach for Polyps Detection in Endoscopic Videos Based on Faster R-CNN}
	\author{\IEEEauthorblockN{Xi Mo\IEEEauthorrefmark{2},
	Ke Tao\IEEEauthorrefmark{3},
	Quan Wang\IEEEauthorrefmark{3} and
	Guanghui Wang\IEEEauthorrefmark{2}}
	\IEEEauthorblockA{\IEEEauthorrefmark{2}Department of Electrical Engineering \& Computer Science, University of Kansas, Lawrence, Kansas 66045, USA}
	\IEEEauthorblockA{\IEEEauthorrefmark{3}the First Hospital of Jilin University, Changchun 130000, China}}
	\maketitle
	\begin{abstract}
		Polyp has long been considered as one of the major etiologies to colorectal cancer which is a fatal disease around the world, thus early detection and recognition of polyps plays an crucial role in clinical routines. Accurate diagnoses of polyps through endoscopes operated by physicians becomes a chanllenging task not only due to the varying expertise of physicians, but also the inherent nature of endoscopic inspections. To facilitate this process, computer-aid techniques that emphasize on fully-conventional image processing and novel machine learning enhanced approaches have been dedicatedly designed for polyp detection in endoscopic videos or images. Among all proposed algorithms, deep learning based methods take the lead in terms of multiple metrics in evolutions for algorithmic performance. In this work, a highly effective model, namely the faster region-based convolutional neural network (Faster R-CNN) is implemented for polyp detection. In comparison with the reported results of the state-of-the-art approaches on polyps detection, extensive experiments demonstrate that the Faster R-CNN achieves very competing results, and it is an efficient approach for clinical practice.
	\end{abstract}
	\section{Introduction}
	\IEEEPARstart{I}{t} is well known that the predecessor of colorectal cancer (CRC), also termed as colon cancer, is most likely to be a polyp. According to the statistics of American Cancer Society, colorectal carcinoma is the third most commonly diagnosed cancer and the second leading cause of death from cancers in the United States~\cite{Wang2015}. CRC is the fourth cause of cancer death worldwide with around 750,000 new cases diagnosed in 2012 alone~\cite{Gil2015}.
	
	Pathologically, neoplastic polyps may chronically turn into cancer, located hiddenly on colorectal wall unless filmed during colonoscopy, which is the main diagnostic procedure of doctors. Though this process may be intuitively achieved successfully, approximately 25\% of polyps are missed~\cite{Tajbakhsh2015}, which brings about potential risks to patients' lives. For early diagnoses and prevention of colon cancer, an urgent task for physicians and computer vision researchers is to find more reliable, accurate and even faster approaches for polyp detection. In response to the demands, well-designed grand challenges organized by Medical Image Computing and Computer Assisted Intervention (MICCAI) and International Symposium on Biomedical Imaging (ISBI), have attracted a lot of attention worldwide.
	
	Specifically, multiple factors affect either the process of manual inspection or computer-aided detection significantly. For the first and foremost, it is common that during the clinical practices, physicians usually operate conventional colonscope for hours to seek, observe, and diagnose polyps, when considering the heavy workload of physicians that leads to both mental and physical fatigue, even an experienced doctor would miss or wrongly diagnose benign polyps. Therefore, automatic computer-aid system is urgently in modern medicine communities~\cite{Albisser2015, Huo2018}.
	
	In regard to computer-aid methods, the factors are diverse. There are varieties of noises in the videos which can be classified as the specular highlights caused by illumination along with the non-Lambertian colorectal walls, the curving veins distributed around the polyps, the polyp like bulges on internal wall to lumen, blob-like matters such as bubbles that always being observed, and the insufficient illumination that shield all regions of interest (ROI). These noises may invalidate the state-of-the-art conventional and learning-based approaches~\cite{Iwahori2015, Tajbakhsh2016, Tajbakhsh2015}. Our experiments show that, in some rare cases, Faster R-CNN~\cite{Ren2015} may mistake some oval specular highlights for polyps.
	
	Another factor is the shape information. Polyps are not always appeared as regular oval lumps, furthermore, they can be various in their sizes from 3\textit{mm} to more than 10\textit{mm} or more variable due to projective transformation and distortions of imaging sensors. Conventional hand-crafted approaches and some fusion approaches often suffer from this factor in that they are initially designed according to the morphological features of polyp~\cite{Wang2015, Tajbakhsh2015, Tajbakhsh2016, Yang2017, Mamonov2014}.\\
	\indent Bernal \textit{et al.}~\cite{Bernal2017} categorize off-the-shelf methods for polyp detection into three classes: hand-crafted, hybrid, and end-to-end learning. Our work emphasizes on the deep learning solution to polyp detection, and provide evaluation of variations in parameters. Our contributions include:
	\begin{itemize}
		\item To our best knowledge, this work provides the first evaluation for polyp detection using Faster-RCNN framework. In addition to reducing the false positive rate during the test phase whose goal is to lower the risks for misdiagnosis when taking the detector for clinical practice, our system provides a good trade-off between efficiency and accuracy.
		\item We demonstrate a fine-tuned set of parameters for polyps detection in endoscopic videos that outperform many state-of-the-art methods. The testing results set a novel baseline for polyp detection.
		\item We compare and analyze the experimental results and reveal insights for better solution when deal with small dataset consisted of endoscopic videos. The proposed framework together with the trained parameters are available for the research community on the author's website.
	\end{itemize}
	\begin{figure}[t]
		\centering
		\subfloat[]{
			\label{fig:4a}
			\includegraphics[width=1.1in]{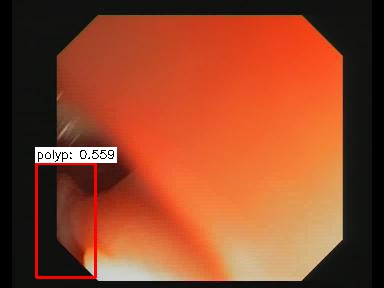}}
		\subfloat[]{
			\label{fig:4b}
			\includegraphics[width=1.1in]{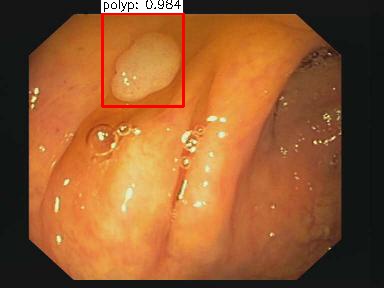}}
		\subfloat[]{
			\label{fig:4c}
			\includegraphics[width=1.1in]{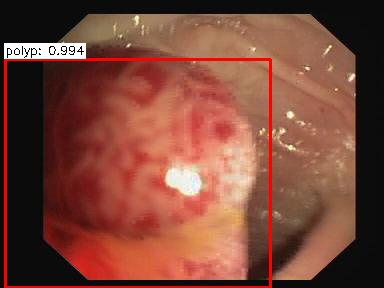}}
		\caption{Features of Faster R-CNN detector. (a) Successfully detect a largely occluded (approximately 50\%) polyp with low intensities. (b) Blob-like objects as bubbles are neglected. (c) Very large polyp detected.}
		\label{fig:4}
	\end{figure}
	\section{Related Works}\label{section:related}
	\begin{figure*}
		\centering
		\includegraphics[width=7.1in]{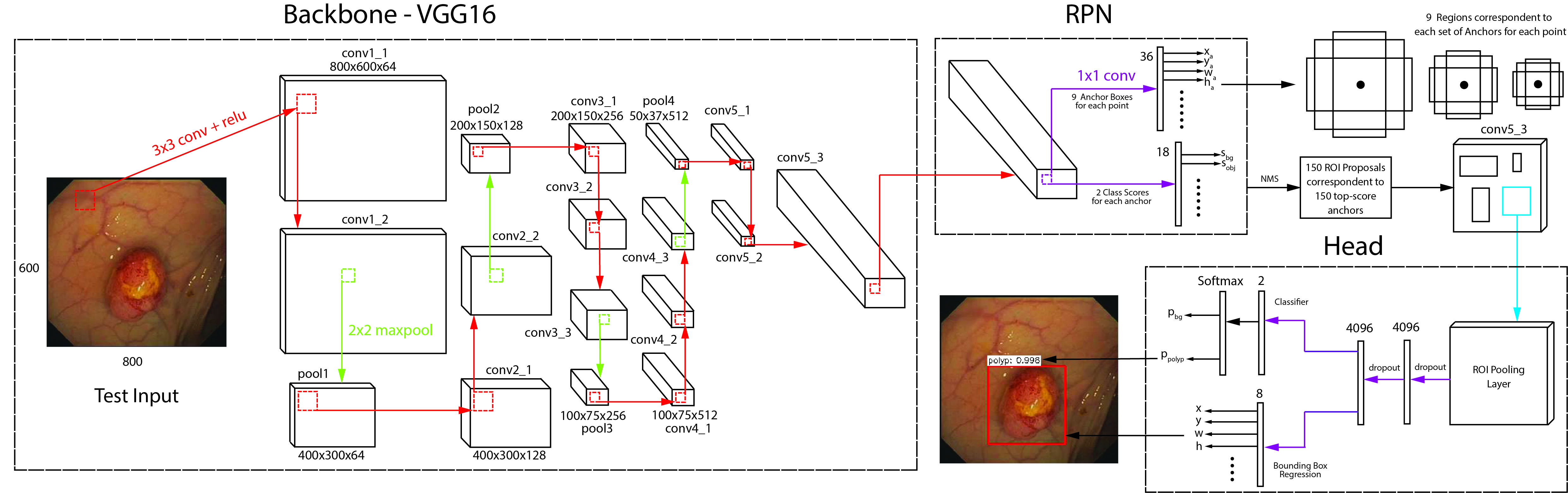}
		\DeclareGraphicsExtensions.
		\caption{Structure for polyp detection. The image is best viewed in its colored version that the red arrow signifies the convolution-ReLU flow, green arrow the max pooling flow, purple arrow the 1$\times$1-convolution or fully-connected flow, blue arrow the ROI pooling flow, and black arrow represents the normal datum flow. Noted that all operations of the same sized convolutional kernel is only labeled at the first appearance of that flow. The input frame is selected from new CVC-ClinicDB (CVC-ClinicDB2017), resizing to 800$\times$600.}
		\label{fig:2}
	\end{figure*}

	According to MICCAI 2015 challenge evaluations, fully CNN based methods with or without data augmentation outperform fusion methods and hand-crafted when considering the evaluation metrics in most cases: Recall, Precision, F-scores (e.g. CUMED, OUS in all videos and videos with only polyp frames). However, high false positive rate has been observed during the experiments~\cite{Roth2017} that a novel data augmentation technique - random view aggregation is implemented, while for pursuing the highest F-scores and remedying the deficiencies of 2D-CNNs, online and offline 3D-fully convolutional networks (FCNs) are integrated to acquire the final confidence map~\cite{Yu2017}. For 2D-CNNs, most of related works focus on no more than 5-convolutional layered deep CNNs such as AlexNet~\cite{Tajbakhsh2016b}, but a few~\cite{Roth2017} have experimented on deeper networks. It remains to be a key topic whether light weighted CNNs can achieve the same capacity as their very deep counterparts. Still, we believe that a trade-off between architectural complexity and runtime would contribute to the ideal design, which is the main reason that we choose VGG16~\cite{Simonyan2014}, once achieved 92.7\% top-5 test accuracy in ImageNet dataset as the feature extractor.
	
	To the best of our knowledge, Faster-RCNN is the first detector so far that replaces hand-crafted ROI selection step with a network i.e., the regional proposal network (RPN) towards fully end-to-end fashion. Its structure is developed from previous R-CNN~\cite{Girshick2014} and Fast-RCNN~\cite{Girshick2015}. Recently, an improvement of Faster R-CNN, i.e., the Mask R-CNN~\cite{He2017} is proposed by extending a novel multi-task branch: mask sub-network for segmentation purpose along with replacement of ROI Pooling layer by ROI Align layer. We apply Faster R-CNN without the sub-network for its redundancy in detecting polyps.
	
	Other novel end-to-end detectors such as You Only Look Once (YOLOv1)~\cite{Khatib2016}, YOLOv2~\cite{Redmon2016a}, SSD~\cite{Liu2016}, so far, most of them have been implemented and tested on other public or private datasets such as COCO, ImageNet, etc. Although these approaches could fulfill realtime requirements (up to more than 24fps), the ROIs are randomly chosen without an end-to-end fashion, and the mAP is compromised in terms of polyp detection as reported in~\cite{Pogorelov2017} that examines YOLOv1 on ASU-Mayo Clinic dataset\cite{Tajbakhsh2016}.
\section{Architecture of Faster R-CNN}\label{section:fasterrcnn}
\subsection{Backbone Structure}
Fig.~\ref{fig:2} illustrates the complete testing structure of this work. The backbone~\cite{He2017} computes high-level features of entire test frame such that the weights between ROIs are shared, which is different from previous R-CNN and patchwise OUS~\cite{Bernal2017} methods. Faster R-CNN removes all subsequent layers of 512 feature maps \textit{conv5\_3} whose shape is 50$\times$37 for each. In reference to VGG16 and ZFnet, it is reported that the latter runs faster up to 17fps, while the former runs at 5fps~\cite{Ren2015}, on a K40 GPU. When comparing mAPs on PascalVOC 2007, ZFnet backbone achieves highest 59.9\%, and VGG16 78.8\%. VGG16 thus benefits for its deep feature extraction process besides its relatively high speed compared to CUMED~\cite{Bernal2017} that runs at 5fps on a more advanced TitanX GPU for former CVC-ClinicDB (CVC-ClinicDB2015)~\cite{Bernal2015}.
	\subsection{RPN and Head Networks}
The $conv5\_3$ is fed to two sibling branches - RPN and Head~\cite{He2017}. After performing 3$\times$3 convolution, RPN constructs 9 anchors at each position on the resulted feature map, the anchors are designed according to 3 scales (small, medium, large) with 3 different ratios of 1:1, 1:2, 2:1. As a result, it outputs maximum 50$\times$37$\times$9$\times$4=66600 positional coordinates of all 16650 potential proposals (for each proposed, the coordinates in the test image are the center $(x_a, y_a)$, width $w_a$ and height $h_a$ of the bounding box), and 50$\times$37$\times$9$\times$2=33300 scores per proposal being the background or polyp. During the training, not all proposals are transformed to training samples, of which a limited number of refined proposals e.g. 2000, are selected by trimming invalided bounding boxes along the borders; proposals with intersection of union (IoU) between 0.3 and 0.7, and in the meantime, keep as many positive samples (IoU$>$0.7) as possible, and replenish with negative samples (IoU$<$0.3); and applying non-maximum suppression (NMS) to the scores $S_{bg}$ and $S_{obj}$, as depicted in Fig.~\ref{fig:2}.

During the testing process, we let RPN generate 150 top proposals further trimmed by NMS of the scores $s_{obj}$ and $s_{bg}$, afterall, RPN is trained for valid regional proposals better than its counterpart - selective search. Refined candidates are then mapped to anchors on $conv5\_3$. The Head network leverages on each anchor to yield the detection outcomes.

As shown in Fig.~\ref{fig:2}, the blue arrow represents the ROI pooling process. All 150 anchors are resized to the same size, which is equivalent to a single-layered SPPnet~\cite{He2014}. This procedure is essential as it transforms different scaled feature map into the two following 4096 fix-length fully-connected layers, each is followed by a dropout layer with a probability of 0.5, which makes the softmax classifier applicable. In addition to regress bounding box of predicted ROIs $(x,y,w,h)$, the Head output 2-class probabilities of the correspondent ROIs to be either background $P_{bg}$ or polyp $P_{polyp}$.
\subsection{Loss Function}
Either RPN or the Head loss functions~\cite{Ren2015} of Faster R-CNN consists of two parts i.e., the classification loss $L_{cls}$ and bounding box regression loss $L_{reg}$. Suppose the ground truth of a proposal to be $\{x^*,y^*,w^*,h^*,P_i^*\}$, among which $P_{bg}^*=1$ and $P_{polyp}^*=0$ if the proposal is positive, and $P_{bg}^*=1$ and $P_{polyp}^*=0$ if negative. To alleviate the influence of scales during training,  the coordinates are parameterized as
	\begin{equation}
	\begin{aligned}
	\left\{
	\begin{array}{ll}
	t_x=(x-x_a)/w_a, & t_y=(y-y_a)/h_a,\\
	t_w=log(w/w_a), & t_h=log(h/h_a),
	\end{array}
	\right.\\
	\left\{
	\begin{array}{ll}
	t_x^*=(x^*-x_a)/w_a, & t_y^*=(y^*-y_a)/h_a,\\
	t_w^*=log(w^*/w_a), & t_h=log(h^*/h_a),
	\end{array}
	\right.
	\end{aligned}
	\end{equation}
	and the general loss function is denoted as
	\begin{equation}\label{eqn:2}
	\begin{aligned}
	&L(\{P_{bg}, P_{polyp}\},\{t_i\})=\frac{1}{N_{cls}}[L_{cls}(P_{bg}, P_{bg}^*)+\\
	&L_{cls}(P_{polyp},P_{polyp}^*)]+\lambda\frac{1}{N_{reg}}\sum_{i}P_i^*L_{reg}(t_i,t_i^*),
	\end{aligned}
	\end{equation}
	where $N_{cls}$ denotes the mini-batch size, $N_{reg}$ the number of all proposals from an image for training. Here the classification loss $L_{cls}(P_i, P_i^*)=-P_i^*log(P_i)$, where $P_{poyp}+P_{bg}=1$, $P_{poyp}$ and $P_{bg}$ are outputs of softmax classifier, and the bounding box regression loss $L_{reg}(t_i,t_i^*)=R(t_i-t_i^*)$, in which $R(\cdot)$ is smooth $L_1$ function for Head loss denoted as
	\begin{equation}
	R(x)=\left\{
	\begin{array}{ll}
	0.5x^2 & |x|<1\\
	|x|-0.5 & otherwise.
	\end{array}
	\right.
	\end{equation}
	
For joint training, the total loss is the sum of RPN and Head losses. while applying 4-step training, two losses are tuned alternately.
	\begin{table*}[h]
		\centering
		\setlength{\belowcaptionskip}{3pt}
		\renewcommand\arraystretch{1.5}
		\caption{Validation metrics.\label{tab:1}}
		\begin{tabular}{c|c|c}
			\hline
			\hline
			& Polyp Detection & Polyp Localization\\
			\hline
			\hline
			True Positive (TP) & Indicate polyp presence in a frame with polyp & Correctly predict polyp location within polyp frame\\
			False Positive (FP) & Indicate polyp presence in a frame without polyp & Wrongly predict polyp location within polyp frame\\
			True Negative (TN) & Indicate polyp missing in a frame without polyp & N/A\\
			False Negative (FN) & Indicate polyp missing in a frame with polyp & Indicate polyp missing in a frame with polyp\\
			\hline
			Precision & $100\times \rm{\frac{TP}{TP+FP}}$ & $100\times \rm{\frac{TP}{TP+FP}}$\\
			Recall & $100\times \rm{\frac{TP}{TP+FN}}$ & $100\times \rm{\frac{TP}{TP+FN}}$\\
			Accuracy & $100\times \rm{\frac{TP+TN}{TP+TN+FP+FN}}$ & N/A\\
			F1-score & $\rm{2\times \frac{Precision\times Recall}{Precision+Recall}}$ & $\rm{2\times\frac{Precision\times Recall}{Precision+Recall}}$\\
			F2-score & $\rm{5\times \frac{Precision\times Recall}{4\times Precision+Recall}}$ & $\rm{5\times \frac{Precision\times Recall}{4\times Precision+Recall}}$\\
			Reaction Time (RT) & Delay between the first TP and polyp frame & N/A\\
			Mean Distance(MD) & N/A & Mean Euclidean distance between polyp centers\\
			\hline
		\end{tabular}
	\end{table*}
	\begin{table*}[h]
		\centering
		\setlength{\belowcaptionskip}{3pt}
		\renewcommand\arraystretch{1.5}
		\caption{Fine-tuned detection results for 300 proposals.\label{tab:2}}
		\begin{tabular}{c|c|c|c|c|c|c|c|c|c|c}
			\hline
			\hline
			Dataset & TP & FP & TN & FN & Accuracy & Precision & Recall & F1-score & F2-score & RT (in frame)\\
			\hline
			\hline
			CVC-Clinic2015-train & 607 & 0 & 0 & 5 & 99.2 & 100.0 & 99.2 & 99.6 & 99.3 & 0\\
			CVC-ColonDB & 292 & 0 & 0 & 8 & 97.3 & 100.0 & 97.3 & 98.6 & 97.9 & 0\\
			CVC-EndoSceneStill & 181 & 0 & 0 & 2 & 98.9 & 100.0 & 98.9 & 99.5 & 99.1 & 0\\
			\hline
			Average & - & - & - & - & \bf{98.5} & \bf{100.0} & \bf{98.5} & \bf{97.1} & \bf{99.2} & \bf{0}\\
			\hline
		\end{tabular}
	\end{table*}
	\begin{table*}[h]
		\centering
		\setlength{\belowcaptionskip}{3pt}
		\renewcommand\arraystretch{1.5}
		\caption{Fine-tuned localization results for 300 proposals and comparison.\label{tab:3}}
		\begin{tabular}{c|c|c|c|c|c|c|c|c|c}
			\hline
			\hline
			Method & Dataset & TP & FP & FN & Precision & Recall & F1-score & F2-score & MD (in pixels)\\
			\hline
			\hline
			\multirow{3}{*}{Faster R-CNN} & CVC-Clinic2015-train & 523 & 81 & 8 & 86.6 & 98.5 & 92.2 & 95.9 & 27\\
			 & CVC-ColonDB & 262 & 30 & 8 & 89.7 & 97.0 & 93.2 & 95.5 & 21\\
			 & CVC-EndoSceneStill & 149 & 32 & 2 & 82.3 & 98.7 & 89.8 & 95.4 & 25\\
			 \hline
			Average & - & - & - & - & \bf{86.2} & \bf{98.1} & \bf{91.7} & \bf{95.6} & \bf{25}\\
			\hline
			Darknet-YOLO-EIR~\cite{Pogorelov2017} & ASU-Mayo Clinic & 2245 & 1005 & 2068 & 69.1 & 52.1 & 59.4 & 62.5 & -\\
			\hline 
		\end{tabular}
	\end{table*}
	\section{Implementation Details}\label{section:details}
	\subsection{Data Preparation}\label{section:preperation}
	The framework is tested using the following public datasets tested during our experiments include:
	\begin{itemize}
		\item \textbf{CVC-Clinic2015 (CVC15)}. Contains 612 still frames whose ground-truths are labeled by the Computer Vision Center (CVC), Barcelona, Spain are selected from 29 endoscopic videos by courtesy of Hospital Clinic, Barcelona, Spain. This dataset is designed as the training set for MICCAI2015 and ISBI2015 sub-challenges for polyp detection in endoscopic videos.
		\item \textbf{CVC-Clinic2017}. A new database for MICCAI2017 endoscopic sub-challenge, which consists of 18 different sequences, and all of which showing no more than one polyp and have up to 11954 frames. The test set contains 18 different videos, and has up to 18733 frames.
		\item \textbf{CVC-ColonDB}~\cite{Bernal2012}. Small public dataset maintained by the CVC group, which contains 300 frames from 15 different videos along with their corresponding ground-truth masks, non-informative region masks, contour of the polyp masks.
		\item \textbf{CVC-EndoSceneStill}~\cite{Vazquez2017}. The CVC group combines CVC-ColonDB with CVC-ClinicDB2015 into a new dataset with explicit divisions for train, test, and validation respectively, which is composed of 912 frames obtained from 44 video sequences collected from 36 patients.
	\end{itemize}

We randomly select 16 sequences from CVC-ClinicDB2017 training set for training Faster R-CNN. To test the performance of trained model on CVC-ColonDB, CVC-ClinicDB2015 and CVC-EndoSceneStill, only the training sets are chosen.
	
Only simple transformations are made to the raw images without augmentation. All training frames are resized to 384$\times$288, which is close to original resolutions of samples for not incorporating much distortions, and in the validation set, monochrome tiff images from CVC-Clinic2015 are transformed to chromatic counterparts. In addition, the training samples are flipped horizontally.
	\begin{figure}[h]
		\centering
		\subfloat[Small polyp.]{
			\label{fig:5a}
			\includegraphics[width=1.68in]{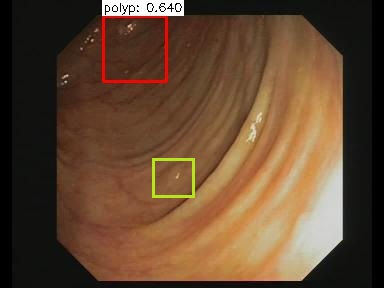}}
		\subfloat[Irrgular shape.]{
			\label{fig:5b}
			\includegraphics[width=1.68in]{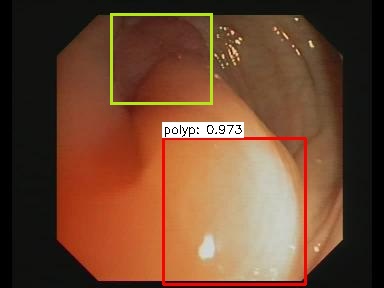}}\\
		\subfloat[Oval specular highlight.]{
			\label{fig:5c}
			\includegraphics[width=1.68in]{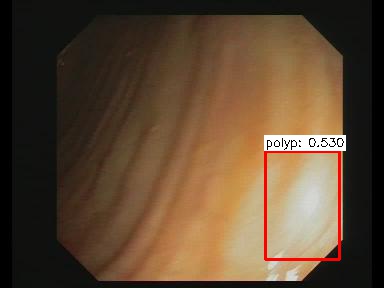}}
		\subfloat[Polyp-like intervention.]{
			\label{fig:5d}
			\includegraphics[width=1.68in]{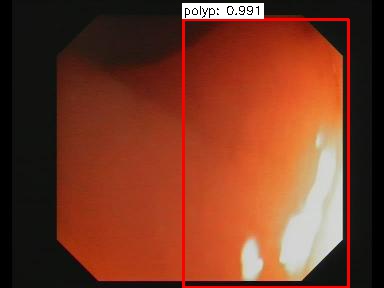}}
		\caption{Failure modes for Faster R-CNN, in which the green bounding box signifies groudtruth. (a) Small-sized polyps are missed. (b) Detector fails to locate low-contrast irregular polyp. (c) Area of specular highlight tricks the detector where there is no polyp. (d) In a frame without polyp, some suspicious area may trick both the detector and human eyes.}
		\label{fig:5}
	\end{figure} 
\subsection{Training}\label{section:train}
Instead of the 4-steps alternately training strategy to optimize RPN and Head losses, we test another approximately joint optimization (AJO) proposed by authors of~\cite{Ren2015} that takes a mini-batch as input and optimizes both losses at the same time. Nevertheless, there is no differential error increments for stochastic gradient descent (SGD) method at RoI pooling layer, the remedy is to propagate these increments backwards without processing. In contrast the 4-steps training methods, AJO has nearly the same test mAP on PascalVOC 2007 whereas faster during training (save up to 9 hours).
	
The training datasets contains 11954 images in total. We train Faster R-CNN on a K40c GPU with default parameters except setting mini-batch size to 128, all batches are normalized by subtraction of fix mean values. Training took no more than 4 days for fine-tuned network without observation of overfitting. In addition, VGG16 is initialized by ImageNet weights. And after 70000 iterations, fully-trained network saw the convergence except for class loss, which indicates that the fully trained Faster-RCNN using AJO may fail to detect polyps. 
\subsection{Validation}\label{section:validation}
Our polyp detection tasks include predicating whether a frame shows a polyp, and localizing the exact location of a polyp. To track training status, we utilize the rest two sequences of CVC-ClinicDB2017 training set as validation sets for evaluating the performance that contain 1178 frames, 910 of which contain a polyp. All evaluation metrics are consistent with MICCAI2017 sub-challenge except F-scores as shown in Tab.~\ref{tab:1}. Noted that FN, TP are counted once per frame, and FP, FN multiple times per frame.

Training sets of other datasets are considered as validation sets except for CVC-EndoSceneStill where the dataset has its own division up to 183 frames. 1, 25, 50, 100, 200, 300 regional proposals are tested respectively for each dataset.
\section{Experiments}\label{section:experiments}
\subsection{Detection}\label{section:detection}
Tab.~\ref{tab:2} show the fined-tuned results of 300 proposals which yields the best performance upon metrics whereas having the longest runtime. Typically, the detector runs at 17fps for 1 proposal which reaches the lower bound of realtime application, and 0.9fps for 300. Parameters are set as follows: Thresholds for RPN NMS, confidence of detection are 0.7 and 0.3, top 1,000,000 proposals before feeding to RPN NMS to ensure 100\% detection rate, a higher confidence threshold 0.5 would drop the rate to 97.4\%. It can be inferred from the detection results that the detection rate reaches a high level for CVC-Clinic2015, CVC-ColonDB and CVC-EndoSceneStill for the reason that each frame of these datasets contains at least one polyp. During the test of the experiments, due to the lower threshold set for confidence, the higher FN rate is observed during detection. Moreover, it is crucial to make a good trade-off between the performance and speed if an automatic detector is designed for real practice. We found on CVC-ColonDB that number of proposals influenced the detection rate greatly that the accuracy reduced from 97.3\% for 300 proposals to 88.3\% for 1 proposal. This trend is identical with that of CVC-Clinic2015 and CVC-EndoSceneStill. 
	\begin{table}[h]
	\centering
	\setlength{\belowcaptionskip}{3pt}
	\renewcommand\arraystretch{1.5}
	\caption{Performance of novel detectors~\cite{Bernal2017} implemented on CVC-Clinic2015DB testing set and Faster R-CNN implemented on training set.\label{tab:4}}
	\begin{tabular}{c|c|c|c|c|c}
		\hline
		\hline
		Method & Dataset & Precision & Recall & F1-score & F2-score\\
		\hline
		\hline
		ASU & CVC15test & 97.2 & 85.2 & 90.8 & 87.4\\
		CUMED & CVC15test & 91.7 & 98.7 & 95.0 & 97.2\\
		CVC-Clinic & CVC15test & 83.5 & 83.1 & 83.3 & 83.2\\
		OUS & CVC15test & 90.4 & 94.4 & 92.3 & 93.6\\
		PLS & CVC15test & 28.7 & 76.1 & 41.6 & 57.2\\
		SNU & CVC15test & 26.8 & 26.4 & 26.6 & 26.5\\
		\bf{Ours} & \bf{CVC15train} & \bf{86.2} & \bfseries{98.1} & \bf{91.7} & \bf{95.6}\\
		\hline
	\end{tabular}
\end{table}
\subsection{Localization}\label{section:local}
Corresponding localization results are shown in Tab.~\ref{tab:3} and~\ref{tab:4}. To compute MD, the Euclidean distance between the center of detected bounding box and that of ground truth is considered as the reference to judge if the detected center locates within the ground truth radius. Relatively high FP, FN rates are observed on CVC-Clinic2017 dataset, while for other three datasets, all metric values lay around 80\%. These results can be regarded as the baseline for polyp localization in future works. In Tab.~\ref{tab:3}, it is denoted that though Darknet YOLO-EIR achieves realtime performance, the metrics are not sufficient for clinic use yet considering our 1 proposal results on CVC-ColonDB that the precision, recall, F1 and F2 scores, MD are 91.3\%, 87.4\%, 89.3\%, 88.1\%, 18 pixels respectively.

In comparison, as is manifested in Tab.~\ref{tab:4}, the outcomes indicate that Faster R-CNN achieves competitive performance compared to novel learning-based techniques, CUMED, ASU, and OUS ~\cite{Bernal2017} on videos with only polyp frames. Noted that these methods take one detection as TP if the detected center falls within the area of ground-truth mask, which is slightly different from MD metric. To be more specific, MD metric implemented here is more strict for it only considers the shortest side of the ground-truth box. During the experiments, we did not validate Faster R-CNN on the private ASU-Mayo Clinic dataset and the MICCAI2015 testing dataset due to their unavailability. However, the design of test set may differ from that of training set, this potential problem is alleviated by the various sets of polyps under different conditions from CVC15 training set and the similar sources of samples.
\subsection{Fine-Tuning vs from Scratch}\label{section:tunescratch}
On small polyp datasets, we are interested in the resultant performances by training from scratch or fine-tuned. For fully trained Faster R-CNN, all weights are initialized by random sampling from Gaussian distribution with zero mean and a standard deviation of 0.01. Fine tuned network manifests high performance during the test as shown in Tab.~\ref{tab:2}-\ref{tab:3}. The fully-trained network, on the other hand, requires a few more days for training, and it has been observed that the lower mAP of fully-trained network might due to the AJO strategy in that the anchors are more sensitive to the initialized weights and RPN fails to provide sufficient positive samples.

The Faster R-CNN detector can detect largely occluded polyp and being robust to illumination changes as is depicted in Fig.~\ref{fig:4a}, also, noises as circular bubbles (Fig.~\ref{fig:4b}) are correctly predicted by the detector, even in the case that there are other tissues except polyp, and the detector correctly localizes the polyp in frames. Another advantage is that very large polyps that may occupy whole receptive field are successfully detected.

On the other aspect, although the detector is more liable to locate large polyps, it misses some very small polyps in the frames, as depicted in Fig.~\ref{fig:5a}, which accounts for the high FP rate in Tab.~\ref{tab:3}, especially when predicting validation sequence 17 of CVC-Clinic2017. It should note that the detector learns the oval shape of polyp so firmly that it mistakes false areas (Fig.~\ref{fig:5b}-\ref{fig:5d}) as the real polyps, which causes high localization FP rate with respect to all datasets. In our future work, we would focus on solutions to these issues.
\section{Conclusion}
Faster R-CNN has been a fully end-to-end approach for object detection tasks on public datasets of natural scenes. For polyp detection and localization in endoscopic videos, this work first applies Faster R-CNN with VGG16 as the backbone. Through extensive experimental evaluation, the proposed approach exhibits potentials for reaching the best performance on precision, as well as yields competitive results in other metrics. The high detection performance indicates that Faster R-CNN could help lower the risk of missing polyps during colonoscopy examination even if RPN predicts only 1 proposal per test. On the other side, Faster R-CNN shows high false-positive rate in frames with presence of polyp during localization tests, which needs to be further investigated and discussed.
\section*{Acknowledgment}
This work was supported in part by the General Research Fund of the University of Kansas under Grant 2228901.
\bibliographystyle{myIEEEtran}
\bibliography{IEEEfull}
\end{document}